# Kinetic Energy-Based Temperature Computation in Non-Equilibrium Molecular Dynamics Simulation

Bin Liu,[1*] Ran Xu,[1] and Xiaoqiao He[2]

[1]*AML, Department of Engineering Mechanics, Tsinghua University, Beijing 100084, China*

[2]*Department of Building and Construction, City University of Hong Kong, Tat Chee Avenue, Kowloon, Hong Kong, China*

[*]Corresponding author: tel: 86-10-6278-6194; fax: 86-10-6278-1824; e-mail:

liubin@tsinghua.edu.cn

## Abstract

The average kinetic energy is widely used to characterize temperature in molecular dynamics (MD) simulation. In this letter, the applicability of three types of average kinetic energy as measures of temperature is investigated, i.e., the total kinetic energy, kinetic energy without the centroid translation part, and thermal disturbance kinetic energy. Our MD simulations indicate that definitions of temperature based on the kinetic energy including rigid translational or rotational motion may yield unrealistic results. In contrast, the thermal disturbance kinetic energy has wider applicability to temperature computation in non-equilibrium molecular dynamics simulation. If small samples need to be used for local temperature, then a calibration approach is proposed to eliminate the sample-size dependence of the average disturbance kinetic energy.

Temperature is one of the fundamental concepts in physics. It is used to measure the hotness or coldness of macroscopic objects. Independent of materials or a thermometer, it represents the intensity of the thermal motion of molecules in microscopic theory. Recently, researchers have found that the definition of temperature strongly influences molecular simulation, especially molecular dynamics (MD), in which the velocities of atoms are continuously adjusted according to various temperature-controlled algorithms. For example, in modeling multi-walled carbon nanotubes as gigahertz oscillators, different oscillation properties, including the quality factor, Q, can be obtained even at the same reference temperature [1-3], which makes the simulation results controversial. In equilibrium statistical mechanics, the absolute temperature is proportional to the average kinetic energy. Rugh [4,5] developed a method to determine the temperature in Hamiltonian dynamical systems, which has been used and promoted by many researchers [6-8]. Nevertheless, it has not yet been ascertained whether all of the components of kinetic energy contribute to temperature in its previous definitions. Many researchers omit the rigid translational kinetic energy from temperature calculation [9], or simply leave out the translation of the mass center during simulation [10]. One question immediately arises, which is addressed in this letter: should the rigid rotational kinetic energy also be omitted from temperature calculation?

Another issue investigated in this letter is the use of the average kinetic energy to compute the temperature of non-equilibrium states in MD simulations, as the definition of absolute temperature out of equilibrium is a much debated topic [8,11-15]. Because many MD simulations aim to model dynamic processes, there is an urgent need to establish a definition of temperature for non-equilibrium states. We believe that a temperature definition should satisfy the following four conditions.

*Condition 1* If a system is in thermodynamic equilibrium, then this definition should yield the same value at different sample points.

*Condition 2* For systems in thermodynamic equilibrium, the temperature from this definition should be almost independent of sample size when this size is sufficiently large.

*Condition 3* The temperature based on this definition should be independent of the choice of reference frame.

*Condition 4* This temperature definition should be reducible to the classical definition in the simple equilibrium situation.

Based on the above conditions, the applicability of various types of average kinetic energy as measurements of temperature is investigated as follows. When the momenta $\mathbf{p}_i$ appear as squared terms in the Hamiltonian, for an *N*-atom system, we can obtain the widely used temperature definition

$$T = \frac{2}{3k_B}\left\langle \frac{1}{N}\sum_{i=1}^{N}\frac{|\mathbf{p}_i|^2}{2m_i}\right\rangle = \frac{2}{3k_B}\langle H\rangle , \qquad (1)$$

where ‹•› denotes the temporal average, $k_B$ is the Boltzmann constant, $m_i$ is the mass of atom $i$, and $H$ is the average kinetic energy for each atom. In classical mechanics of discrete systems, the atom velocity can be decomposed into three parts as

$$\mathbf{v}_i = \mathbf{v}_\mathbf{C} + \boldsymbol{\omega}\times\hat{\mathbf{r}}^i + \mathbf{v}_\mathbf{i}^{Dis} , \qquad (2)$$

where $\hat{\mathbf{r}}^i$ is the atom position vector relative to the center of mass, $\mathbf{v}_\mathbf{C}$ is the centroid translational velocity, $\boldsymbol{\omega}$ is the rotational velocity around the centroid, and $\mathbf{v}_\mathbf{i}^{Dis}$ is the disturbance velocity. Correspondingly, there are three possible types of average kinetic energy, i.e.,

$$H^{Total}=\frac{1}{N}\sum_{i=1}^{N}\frac{1}{2}m_i\mathbf{v}_i\cdot\mathbf{v}_i=\frac{1}{2N}\left(\sum_{i=1}^{N}m_i\mathbf{v}_i^{Dis}\cdot\mathbf{v}_i^{Dis}+\boldsymbol{\omega}\cdot\mathbf{J}_C\cdot\boldsymbol{\omega}+M\mathbf{v}_C\cdot\mathbf{v}_C\right), \quad (3)$$

$$H^{Dis+R}=\frac{1}{2N}\left(\sum_{i=1}^{N}m_i\mathbf{v}_i^{Dis}\cdot\mathbf{v}_i^{Dis}+\boldsymbol{\omega}\cdot\mathbf{J}_C\cdot\boldsymbol{\omega}\right), \quad (4)$$

and

$$H^{Dis}=\frac{1}{2N}\sum_{i=1}^{N}m_i\mathbf{v}_i^{Dis}\cdot\mathbf{v}_i^{Dis}, \quad (5)$$

where $M$ is the total mass of the system, and $J_C$ is the rotational inertia. Obviously, $H^{Total}$, $H^{Dis+R}$, and $H^{Dis}$ are the averages of the total kinetic energy, kinetic energy without the centroid translation part, and thermal disturbance kinetic energy, respectively. It should be pointed out that $H^{Dis}$ is independent of the reference frame, and is the minimum of $H^{Total}$ for all possible reference frames; therefore, it may serve as an objective quantity to characterize temperature.

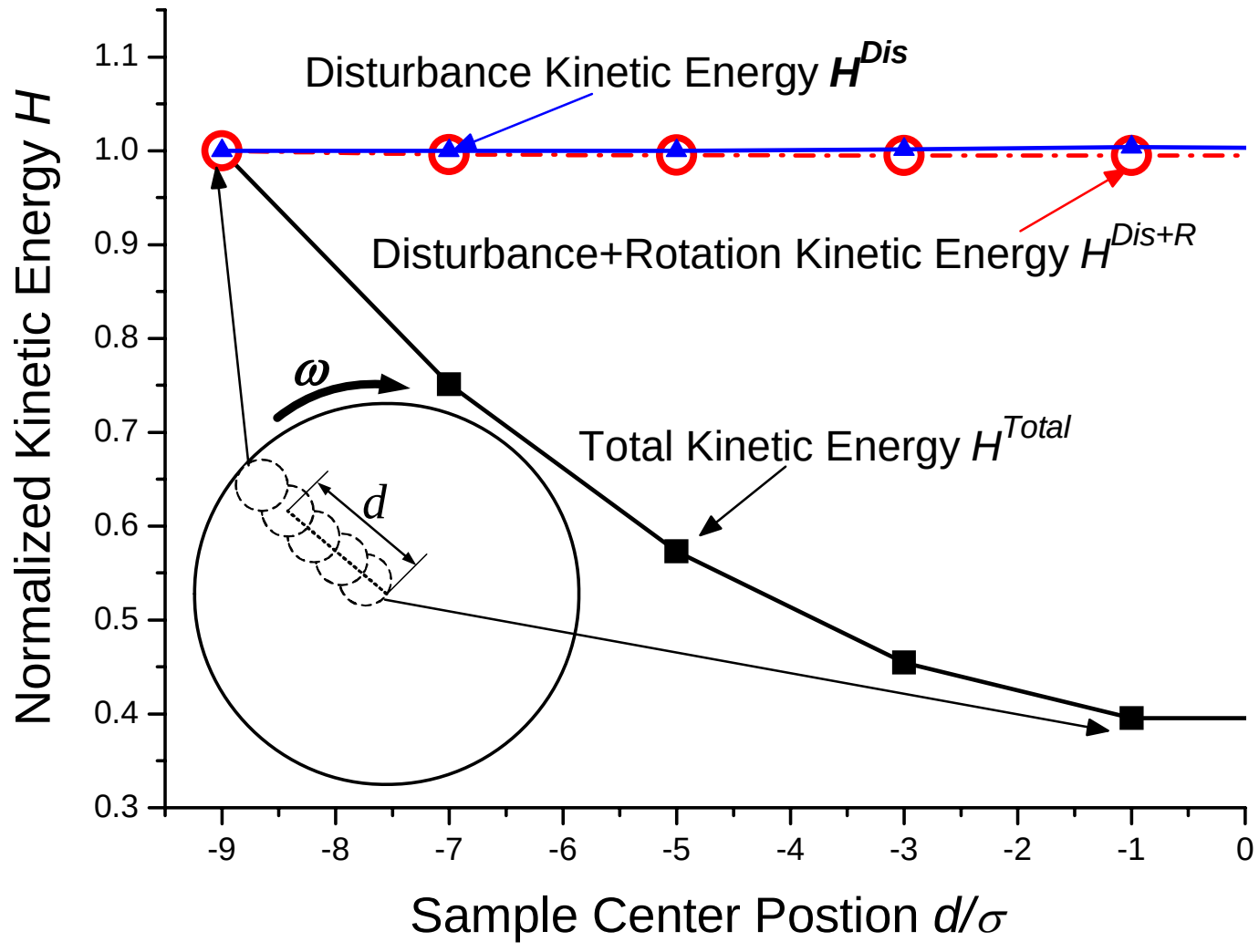

Fig. 1. The normalized average kinetic energy as a function of the sample point position in a rotational atomic system in thermodynamic equilibrium.

To test the definitions of temperature based on various kinds of kinetic energy, we first study a rotational atomic ball with 8630 atoms, which is schematically shown in the inset of Fig. 1, and the interatomic potential is the Lennard-Jones potential $V(\mathbf{r})=4\varepsilon\left(\left(\frac{\sigma}{r}\right)^{12}-\left(\frac{\sigma}{r}\right)^{6}\right)$, where $\varepsilon$ is the depth of the potential well, $\sigma$ is the distance at which the interatomic potential is zero, and $r$ is the distance between the atoms. The rotational ball is simulated as an isolated system without external pressure. Initially, each atom has a thermal random velocity and the system has a rotational velocity. After a sufficient amount of time has passed, this isolated system finally reaches a state of thermodynamic equilibrium, in which all physical state variables remain statistically unchanged over time but the rotational momentum is non-zero and conserved. For this system in thermodynamic equilibrium, the temperature should be uniform, as stated earlier in *Condition 1*.

Normalized by the corresponding outmost values, the three types of average kinetic energy from different sample points of the rotational ball are shown in Fig. 1, and each sample consists of 860 neighbor atoms. It is found that the two average kinetic energies excluding the translational term, $H^{Dis}$ and $H^{Dis+R}$, are uniform over all sample points. However, the average total kinetic energy, $H^{Total}$, has significantly different values for different sample points, which indicates that the average total kinetic energy should not be used as the measurement of temperature in some cases.

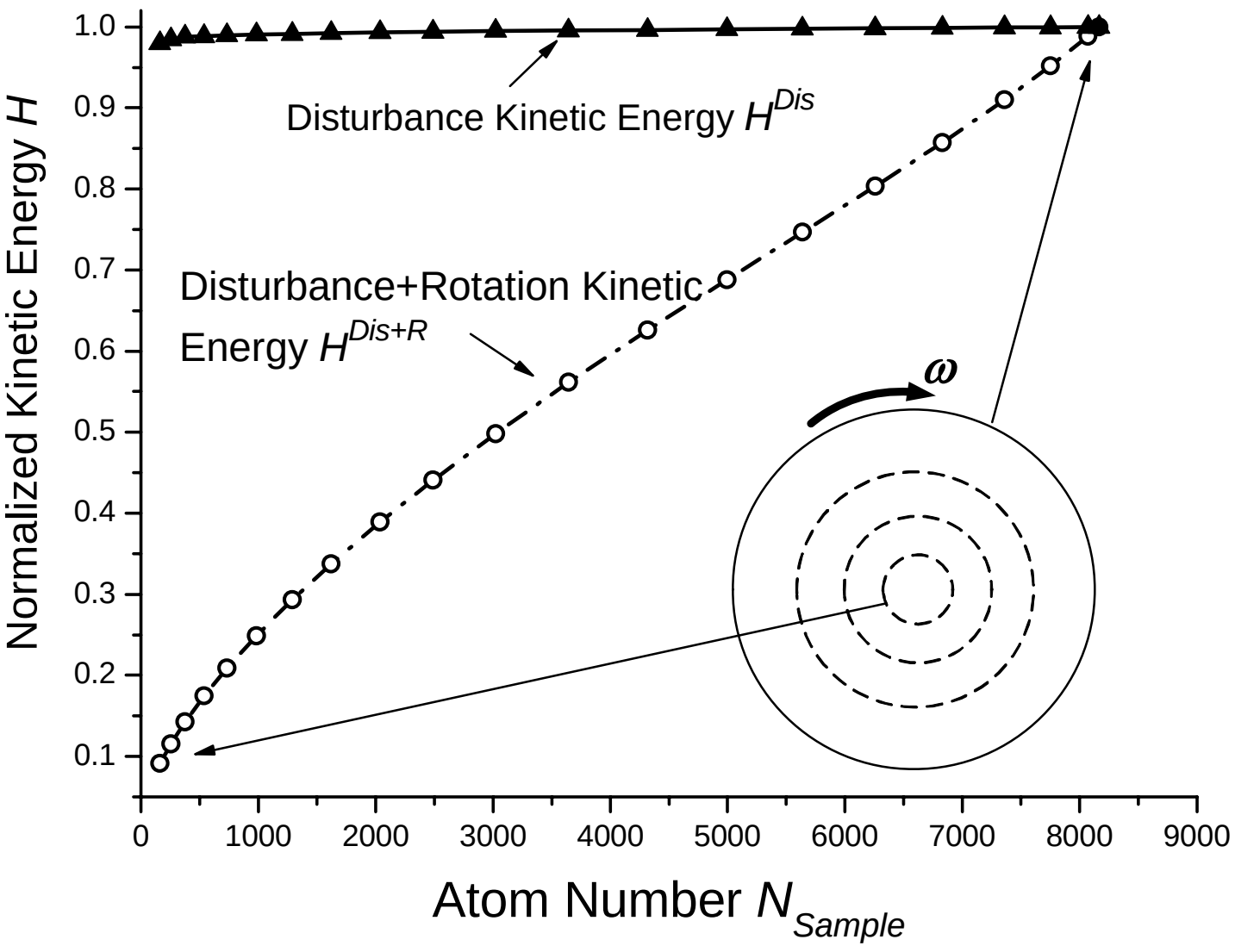

Fig. 2. The normalized average kinetic energy at the center as a function of the sample size.

To investigate whether *Condition 2* is satisfied, we study the effect of sample size on the values of the average kinetic energies $H^{Dis}$ and $H^{Dis+R}$ using this rotational ball example. As shown in Fig. 2, different-sized concentric sphere samples are tested. The MD simulation results indicate that with an increase in the sample size, the average disturbance kinetic energy, $H^{Dis}$, quickly converges to a constant value, whereas $H^{Dis+R}$ does not converge and continuously increases, hence violating *Condition 2*.

It is obvious that as temperature measures, $H^{Total}$ and $H^{Dis+R}$ violate *Condition 3*, because different reference frames will lead to different $H^{Total}$ and $H^{Dis+R}$. In contrast, the average disturbance kinetic energy, $H^{Dis}$, is independent of the choice of reference frame, and thus satisfies *Conditions 1*, *2*, and *3*; therefore, it may be an appropriate candidate to characterize temperature.

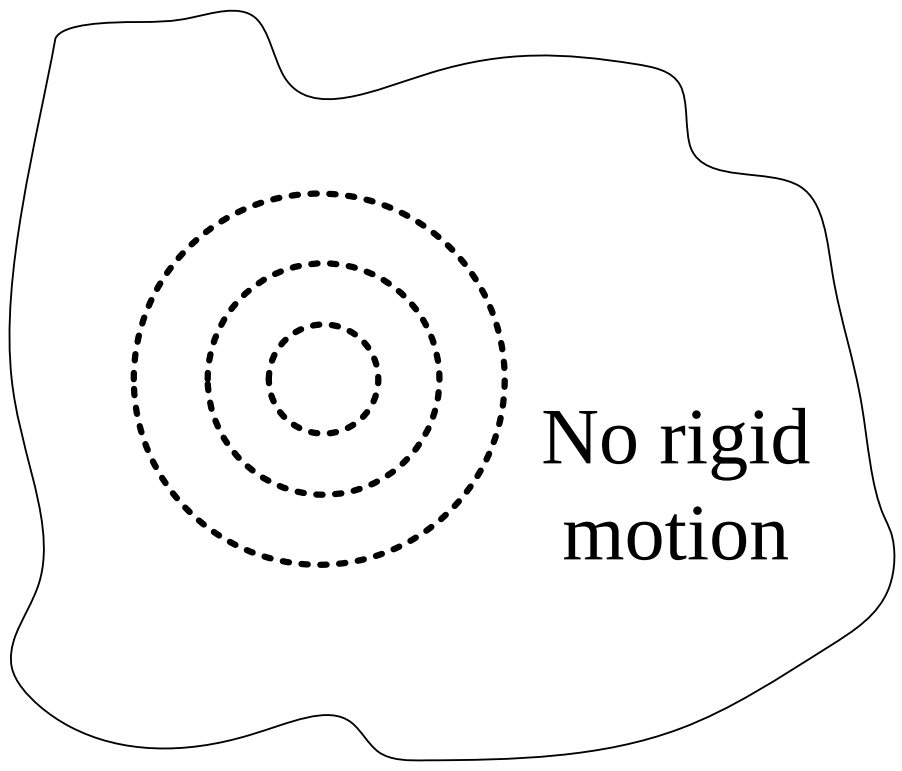

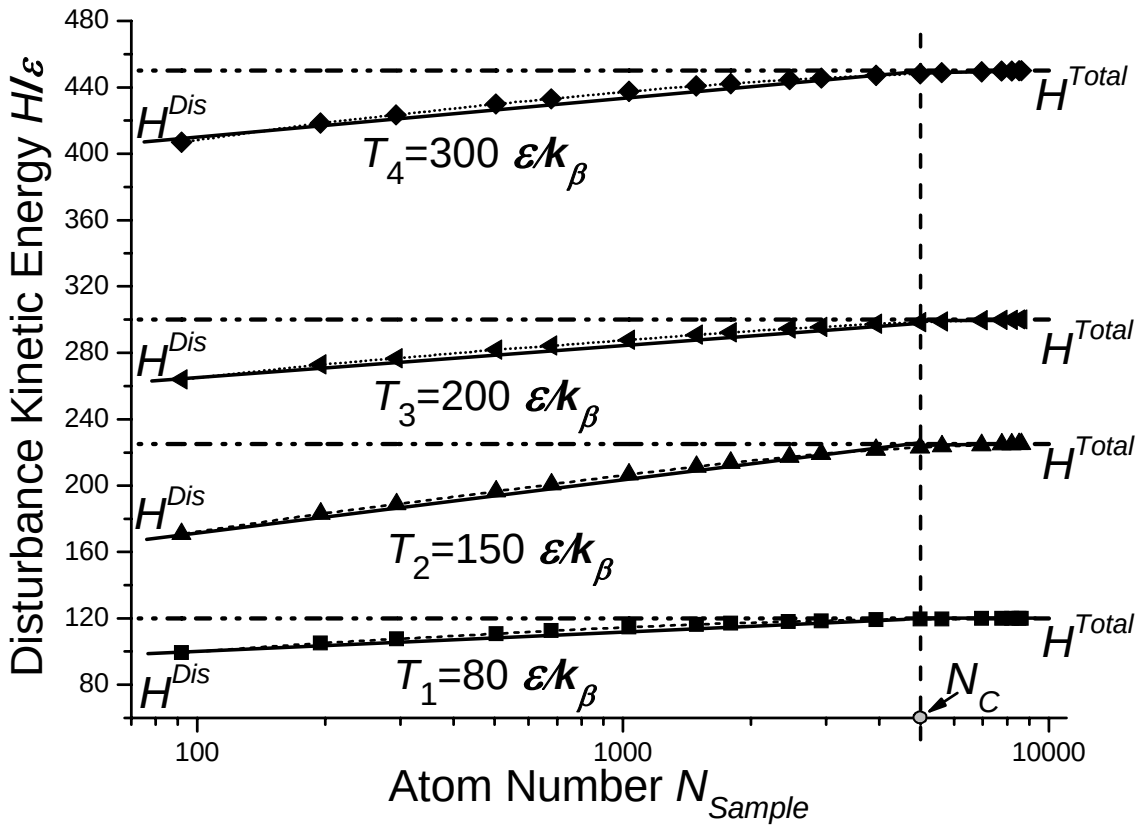

(a) (b)

Fig. 3. (a) Schematic diagram of a system in simple equilibrium without rigid motion; (b) the average disturbance kinetic energy, $H^{Dis}$, as a function of sample size.

To test whether $H^{Dis}$ satisfies *Condition 4*, the system in simple equilibrium without rigid motion shown in Fig. 3(a) is investigated. In this simple equilibrium case, the classical definition of temperature, $T = \frac{2}{3k_B}\left\langle H^{Total} \right\rangle$, is widely accepted (of course, this definition may fail for other complex situations, as discussed earlier), and can serve as a benchmark. It is found that $H^{Dis}$ converges to $H^{Total}$ for a sufficiently large sample. Therefore, the proposed definition of temperature, $T = \frac{2}{3k_B}\left\langle H^{Dis} \right\rangle$, with a large sample size is reducible to the classical definition, which satisfies *Condition 4*.

For a smaller sample, $H^{Dis}$ is dependent on the sample size, as shown in Fig. 3(b), and $H^{Dis}$ should be denoted as $H^{Dis}\left(N_{sample}\right)$, where $N_{sample}$ is the number of atoms in a sample. In some situations, such as non-equilibrium, smaller samples have to be used in MD to obtain the local temperature. In this case, we suggest that the above system in simple equilibrium can be used as a thermometer, based on the assumption that the temperatures of two samples are the same if their sample size ($N_{sample}$) and average disturbance kinetic energy ($H^{Dis}$) are identical. For example, Fig. 3(b) is a temperature contour map obtained from a system in equilibrium without rigid motion, and can serve as a calibration chart to predict the temperature for the non-equilibrium state. If the average disturbance kinetic energy $H^{Dis}\left(N_{sample}\right)$ and the sample size, $N_{sample}$, are known, then a proper temperature can be obtained by interpolation in Fig. 3(b).

Phenomenological scaling laws that can simplify this calibration process may exist. When $N_{sample}$ is plotted along the logarithmic horizontal coordinate, it is noted from Fig. 3 that the curve for $H^{Dis}$ can be approximated as a straight line if $N_{sample}$ is smaller than a critical value, $N_C$. Beyond $N_C$, $H^{Dis}$ converges to $H_{\infty}^{Dis}$ and the curve becomes an approximately horizontal line. More interestingly, a common turning point, $N_C$, is observed for the system with different temperatures. Therefore, $H_{\infty}^{Dis}$ can be computed with the two smaller sized samples $N_1^{Sample}$, $N_2^{Sample}$ and their corresponding average disturbance kinetic energies $H_1^{Dis}$ and $H_2^{Dis}$ as

$$H_{\infty}^{Dis} = H_1^{Dis} + \left(H_2^{Dis} - H_1^{Dis}\right)\frac{\ln\left(N_C / N_1^{Sample}\right)}{\ln\left(N_2^{Sample} / N_1^{Sample}\right)}. \quad (6)$$

In addition, the temperature $T = \frac{2}{3k_B}\left\langle H_{\infty}^{Dis}\right\rangle$ is obtained.

Satisfying all conditions, the definition of temperature based on the average disturbance kinetic energy, $T = \frac{2}{3k_B}\left\langle H^{Dis}\right\rangle$, has wide applicability, as the following dynamic example demonstrates. We simulate the dynamic behavior of an atomic bar, shown in the inset of Fig. 4, which includes 9850 atoms with an harmonic interatomic potential $\frac{1}{2}k\left(r_{ij} - b\right)^2$, where $k$ is the harmonic force constant and $b_0$ is the equilibrium bond length. The bar is initially under uniaxial static tension, and is in thermomechanical equilibrium at $T = 0.15\,kb^2/k_\beta$. The loading is then suddenly released, and the bar starts to vibrate. It can be expected that this system will experience quasi-harmonic vibration, at least in the first several cycles, so many characterized physical quantities will exhibit a sine curve with respect to time. Figure 4 shows the total kinetic energy of the whole system as a function of the normalized time for the three energy/temperature control modes. If this atomic bar is simulated as an isolated system in our MD simulation, i.e., the total energy (including the kinetic and potential parts) is conserved, then the corresponding curve is really a sine curve as denoted by the solid line. In the other two cases, the bar is divided into 15 samples, and the Nose-Hoover [17,18] temperature control method is used to simulate a constant temperature environment by adjusting the average kinetic energy in each sample. If the average total kinetic energy, $H^{Total}$, is adopted in the thermostat, then the curve denoted by the dashed line in Fig. 4 deviates significantly from the sine curve, which seems unrealistic and unreasonable. However, the simulation with the thermostat based on the average disturbance kinetic energy, $H^{Dis}$, yields a dotted sine curve almost consistent with the isolated one. Therefore, the $H^{Dis}$-based definition of temperature is suitable for more complex situations in MD simulations.

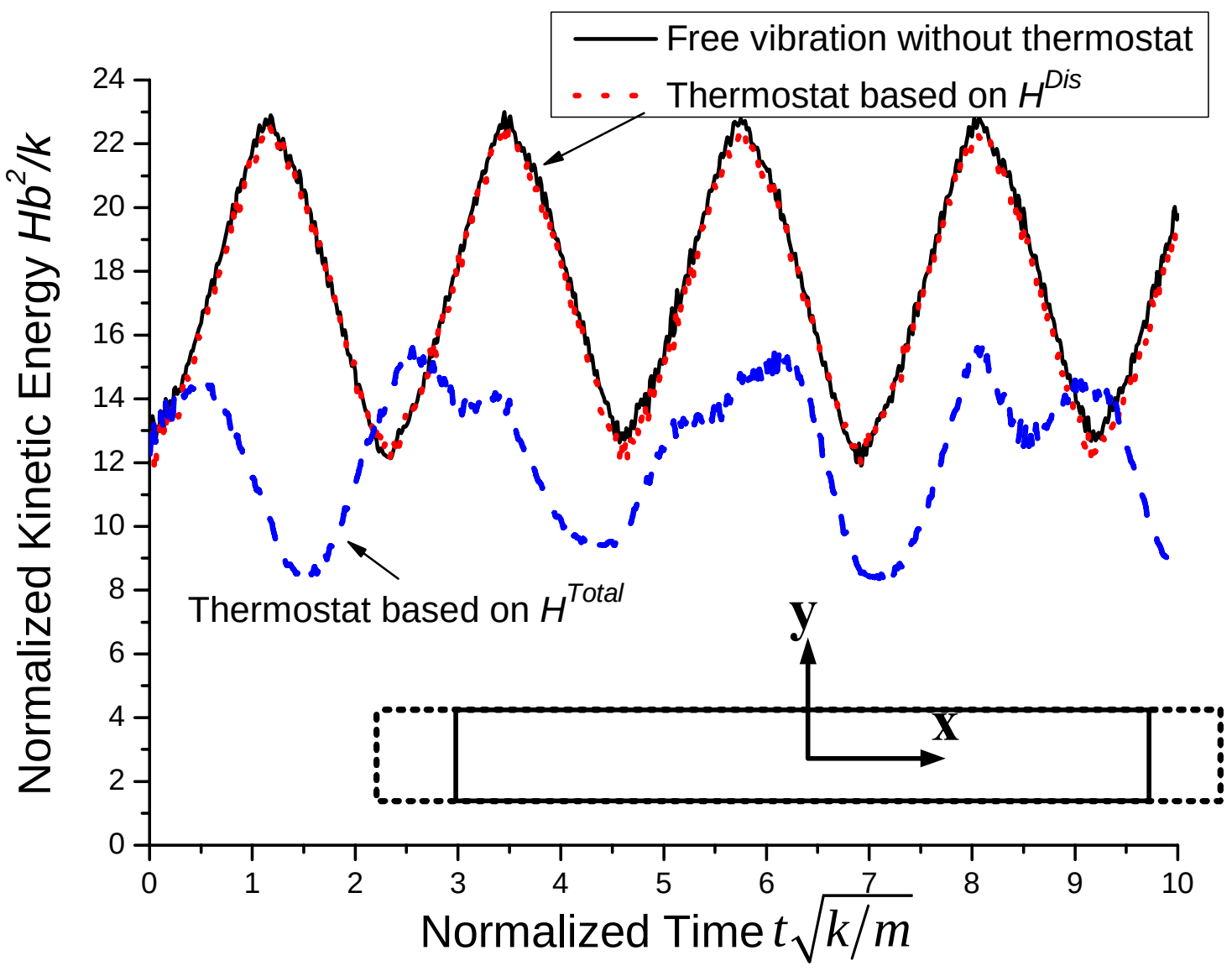

Fig. 4. The normalized kinetic energy of a bar under vibration versus the normalized time.

## Acknowledgements

The authors acknowledge the support from the National Natural Science Foundation of China (grant nos. 10702034, 10732050, 90816006, and 10820101048) and National Basic Research Program of China (973 Program, grants no. 2007CB936803 and 2010CB832701).